\input epsf
\documentstyle[preprint,eqsecnum,aps]{revtex}

\begin{document}

\count255=\time\divide\count255 by 60 \xdef\hourmin{\number\count255}
  \multiply\count255 by-60\advance\count255 by\time
 \xdef\hourmin{\hourmin:\ifnum\count255<10 0\fi\the\count255}

\preprint{\hbox{WM-99-112}}

\title{Electroweak Constraints on Extended Models with Extra Dimensions}

\author{Christopher D. Carone}

\vskip 0.1in

\address{Nuclear and Particle Theory Group, Department of
Physics, College of William and Mary, Williamsburg, VA 23187-8795}

\vskip .1in
\date{July, 1999}
\vskip .1in

\maketitle
\tightenlines
\thispagestyle{empty}

\begin{abstract}
Electroweak measurements place significant bounds on higher-dimensional 
versions of the standard model in which the gauge and Higgs fields 
have Kaluza-Klein excitations.  These bounds may be altered quantitatively 
if chiral matter is also allowed to propagate in the higher-dimensional 
`bulk'.  We determine the electroweak constraints on a number of models 
of this type, including scenarios in which only the leptons or only 
the first two generations of matter fields propagate in the bulk.  We also 
consider the possibility that different factors of the electroweak gauge 
group may be distinguished by their bulk/three-brane assignment, and 
study a minimal extra-dimensional $Z'$ model.  We find typical bounds 
on the compactification scale between $1.5$ and $4$~TeV, and comment
on models in which these bounds might be significantly relaxed.
\end{abstract}


\newpage
\setcounter{page}{1}

\section{Introduction}

The possibility that nature may reveal the presence of extra spacetime
dimensions at distance scales as large as an inverse 
TeV~\cite{xdtev1,xdtev2} has fueled considerable interest in 
extra-dimensional embeddings of the standard 
model~\cite{xdsm1,xdsm2,xdsm3}.  In the minimal approach, chiral 
matter fields are confined to three spatial dimensions, or a 
three-brane, while the gauge and one or both Higgs fields are allowed 
to propagate in the higher, $4+\delta$ dimensional bulk spacetime, 
where $\delta$ dimensions are compactified on an orbifold of 
radius $R$\footnote{There is also the possibility of large extra
gravitational dimensions, which we do not consider here.
See, for example, Ref.~\cite{add}.}.  This picture has many desirable 
features, including the possibility of understanding the breaking of 
supersymmetry at low energies (as well as the origin of the $\mu$ parameter) 
via the Scherk-Schwartz mechanism~\cite{xdsm1,xdsm2,xdss}, and the potential 
of achieving an accelerated gauge 
unification~\cite{xdsm3,kakn,carone1,xdgu1,xdgu2,xdgu3,xdgu4,xdgu5,kak}.  
Bounds on the scale of compactification have been determined in effective 
four-dimensional theories through the effects of Kaluza-Klein (KK) 
excitations on precisely measured low-energy electroweak observables, 
and found to be typically of order a few 
TeV~\cite{ewbds1,ewbds2,ewbds3,ewbds4}.  Hence, the possibility 
exists that KK excitations of the standard model gauge fields might 
be produced and studied at a range of future colliders 
experiments~\cite{ewbds3,kksigs1,kksigs2}.

Within this framework, it is possible to construct models in which 
matter that is chiral under the standard model gauge group also propagates 
in the higher dimensional bulk.  This can be arranged if chiral conjugate 
mirror fields are introduced so that KK mass terms can be formed.  In 
the $Z_2$ orbifold models of interest to us in this paper, these mirror fields 
are taken to be $Z_2$ odd, so that they have no effect on the spectrum of 
light states.  One interesting feature of these models is that a coupling
between any number of bulk fields respects a conservation of KK number.
For example, a coupling between three $Z_2$ even fields 
$\phi_1\phi_2\phi_3$, with $\phi_i = \sum_{n=0} \phi_i^{(n)}\cos(n x_5 /  R)$
in the case of one extra dimension $x_5$, leads to a coupling between
the different modes
\begin{equation}
{\cal L}_{4D}=c_{ijk} \phi^{(i)}_1 \phi^{(j)}_2 \phi^{(j)}_3  
\,\,\, \mbox{ with } \,\,\,
c_{ijk} = \int dx_5 \cos(\frac{ix_5}{R}) \cos(\frac{jx_5}{R}) 
\cos(\frac{kx_5}{R})
\end{equation}
that vanishes if, for example, $i=j=0$ and $k\neq 0$.  Clearly,
models with chiral matter in the bulk will exhibit a different pattern
of couplings between ordinary `zero-mode' particles and the KK excitations 
of the gauge fields.  This suggests that bounds from precision electroweak 
measurements will be affected and therefore merit a reexamination in 
this context.

Which extended models should we consider?  While there are admittedly a 
large number of possibilities, we will focus on three cases that 
are plausible from the point of view of simplicity, and that also 
have potentially interesting low-energy phenomenologies.  The first model,
the `bulk lepton' scenario \cite{carone1}, is somewhat similar in spirit 
to the `ununified' standard model \cite{unun}, though unification 
is in fact one of its strong points.  In the ununified standard model, 
quarks and leptons are distinguished at short distance scales by having 
completely independent electroweak gauge groups.  In the bulk lepton 
scenario, quarks and leptons are distinguished at short distance scales 
by the fact that only leptons can propagate into the higher dimensional 
space.  As we point out in Section~\ref{sec:two}, this model leads to an 
improvement in accelerated gauge unification, and thus presents a 
plausible alternative to the minimal approach.  The second case that we 
consider, the `bulk generations' scenario, has similarities in spirit 
to topcolor~\cite{topmod1} or topflavor models~\cite{topmod2}.  In these
models, the third generation is distinguished from the first two at short 
distance scales by having its own independent set of standard-model-like 
gauge factors.  In the bulk generations scenario, the third generation 
is distinguished at short distances by the fact that it is the only 
generation that cannot propagate into the bulk. We will describe later why 
this selection of bulk generations is preferred for TeV-scale 
compactifications.  In both the bulk lepton and bulk generations 
scenarios, the KK excitations of the standard model gauge fields cannot 
couple to leptons of the light generations, so one might suspect that the 
form of the corrections to electroweak observables would be significantly 
affected.  The third case that we consider is one in which the SU(2)
gauge multiplet is confined to the three-brane, while the others are
not.  As far as the electroweak sector of this model is concerned, one
might make the comparison to four-dimensional models in which a $Z'$ boson 
is obtained by minimally extending the standard model gauge group by a U(1) 
factor.  In the `SU(2)-brane' scenario, a $Z'$ boson is obtained by 
minimally extending the electroweak gauge group into extra dimensions, 
allowing only hypercharge to propagate in the bulk.  One might regard 
the lightest KK state as as a minimal extra-dimensional $Z'$, $Z'_{XD}$, 
that could be placed among many others ($Z'_\eta$, $Z'_\chi$, $Z'_{SM}$, 
$Z'_{LR}$, etc.) that have been studied in the literature.  We consider 
the electroweak bounds on each of these scenarios in 
Sections~\ref{sec:two}, \ref{sec:three}, and \ref{sec:four}, respectively. 
In the final section we summarize our conclusions.

Before proceeding to the analysis, however, it is important to point out 
that we will determine bounds on the compactification scale, $M_c = 1/R$, 
only in the five dimensional version of each of the models described above.  
With only one additional spatial dimension, sums over intermediate KK 
states are  convergent, and results can be obtained without ambiguity.  For
more extra dimensions, the sums over KK states diverge if the gauge
couplings are naively assumed to be independent of KK number.  In 
reality, one expects that there is a suppression of the couplings to higher
KK modes, and that this effect physically regulates the sums~\cite{kksigs1}.  
The dependence of the couplings on KK number, $g(|\vec{n}|^2)$, follows from 
string theory considerations, and is model dependent.  Thus, we will 
relegate a quantitative treatment of the $\delta > 1$ models to a time when 
this dependence is more reliably known.

\section{Bulk Leptons} \label{sec:two}

The possibility that we consider in this section is Scenario~1 of
Ref.~\cite{carone1}: The standard model gauge multiplets as well
as leptons of all three generations live in the higher dimensional
bulk, while the quarks and both Higgs fields are confined to the 
three-brane.  This choice leads to a marked improvement in gauge
unification compared to the minimal scenarios discussed extensively 
in the literature.   In addition, this ``bulk lepton scenario'' predicts 
that the KK excitations of the $W$ and $Z$ bosons will be leptophobic, 
as a consequence of the conservation of KK number.  The fact that
this scenario is a  viable alternative to the minimal one, and
may yield relatively exotic collider signatures (such as a leptophobic 
$W'$ with otherwise standard model couplings) is motivation for 
considering the indirect constraints on the model. It is worth 
mentioning that the choice of placing both Higgs fields on the 
three-brane implies that the $\mu$ parameter is not generated through 
compactification as, for example, in Ref.~\cite{xdsm1}.  However, there 
may be other natural ways of obtaining a $\mu$ parameter of the desired 
magnitude.  We comment on this issue at the end of this section.

The essential phenomenological features of this model can be appreciated 
by considering first a five-dimensional U(1) gauge theory, spontaneously 
broken by two Higgs fields both confined to a three-brane.  The relevant 
four-dimensional Lagrangian is given by
\[
{\cal L}_4 = \sum_{n=0} \left[-\frac{1}{4} F^{\mu\nu (n)} F_{\mu\nu}^{(n)}
+\frac{1}{2} \frac{n^2}{R^2} A^{(n)}_\mu A^{\mu(n)}\right]
+ g^2(v_1^2+v_2^2)(A_\mu^{(0)}+\sqrt{2}\sum_{n=1}A_\mu^{(n)})^2
\]\begin{equation}
+ i \overline{q}( \sigma^\mu\partial_\mu + i g A_\mu^{(0)}
+ i g \sqrt{2} \sum_{n=1} A_\mu^{(n)} ) q
\end{equation}
where $q$ represents any quark field, and the $v_i$ are Higgs vacuum 
expectation values (vevs).  The fact that both Higgs fields 
are stuck on a brane leads to the mixing term between the zero-mode
gauge field and its KK excitations.  The gauge boson mass matrix
takes the form
\begin{equation}
\left(\begin{array}{cccc}
m_z^2 & \sqrt{2} m_z^2 & \sqrt{2} m_z^2 & \cdots \\
\sqrt{2} m_z^2 & M_c^2 &  &  \\
\sqrt{2} m_z^2 &       & (2M_c)^2 & \\
\vdots & & & \ddots \\
\end{array}\right)
\label{eq:massmat}
\end{equation}
where $m_z^2=2g^2(v_1^2+v_2^2)$.   Working to lowest order in 
$m_z^2/M_c^2$, it is straightforward to show that this matrix
is diagonalized by the rotation
\begin{equation}
R=
\left(\begin{array}{cccc}
1 & \theta_1 & \theta_2 & \cdots \\
-\theta_1 & 1 & & \\
-\theta_2& & 1 & \\
\vdots &   &  & \ddots \\
\end{array}\right)   \,\,\,\,\,\,\,\,\,\, 
\theta_n=-\frac{\sqrt{2}m_z^2}{n^2 M_c^2}
\label{eq:rmat}
\end{equation}
and that the lowest eigenvalue is given by
\begin{equation}
m_z^{(ph)2} = m_z^2 (1-2 \sum_{n=1} \frac{m_z^2}{n^2 M_c^2}) \,\,\, .
\label{eq:mzormw}
\end{equation}
Unlike the case in which one Higgs field lives in the bulk~\cite{xdsm1}, 
here there is no dependence on the ratio of the Higgs vacuum expectation 
values (vevs).  The rotation in Eq.~(\ref{eq:rmat}) leads to a shift 
in the coupling of the zero-mode gauge field to zero-mode brane fermions 
(in this case the quarks and Higgs)
\begin{equation}
g^{(ph)}_{brane} = g (1-2\sum_{n=1} \frac{m_z^2}{n^2 M_c^2}) \,\,\, ,
\label{eq:gphbrane}
\end{equation}
but no change in the coupling to zero-mode bulk fermions 
(the leptons)
\begin{equation}
g^{(ph)}_{bulk} = g  \,\,\, .
\end{equation}
The generalization of these results to an SU(2)$\times$U(1)
gauge theory is straightforward: one breaks electroweak symmetry
in the five dimensional theory and rotates from the weak to mass
eigenstate basis before integrating over $x^5$.  Then one obtains
mass matrices for the $W$ and $Z$ bosons that are of the same form 
as Eq.~(\ref{eq:massmat}), with $m_z \rightarrow m_Z\mbox{,}m_W$,
respectively.  With these results in hand, we may now consider the 
corrections to electroweak observables.  We will denote the SU(2)
and U(1) gauge couplings by $g$ and $g'$, respectively.  For clarity,
we will express our analytic results in terms of sums over
KK modes.  To obtain numerical results, we will use 
$\sum_{n=1} 1/n^2=\pi^2/6$ for $\delta=1$.

Let us begin with the Fermi constant.  The KK excitations of the $W$ 
do not contribute to muon decay at order $m_W^2/M_c^2$.  However, $G_F$ 
is still affected by the shift of the $W$ and $Z$ mass eigenvalues.  
Given the standard model relation,
\begin{equation}
G_F^{SM}= \frac{\pi \alpha}{\sqrt{2} m_W^{(ph)2} \left(
1-\frac{m_W^{(ph)2}}{m_Z^{(ph)2}}\right)(1-\Delta r)} \,\,\, ,
\end{equation}
we find
\begin{equation}
G_F = G_F^{SM}\left[1-2\sum_{n=1} \frac{m_W^{(ph)2}}{n^2 M_c^2}
\right] \,\,\, .
\label{eq:fermi}
\end{equation}
Note that the standard model radiative corrections are subsumed into
$\Delta r$, so that the expression in (\ref{eq:fermi}) is accurate to 
order $\Delta r$, but not order $\Delta r (m_W^2/M_c^2)$.  Using the 
experimental value 
$G_f=1.16639\pm 0.00001\times 10^{-5}$~GeV$^{-2}$~\cite{rpp98}, 
and $G_F^{SM}=1.16775\pm 0.0049 \times 10^{-5}$~GeV$^{-2}$ computed from 
the $W$ and $Z$ masses~\cite{ewbds1}, we obtain the bound
\begin{equation}
M_c > 1.49 \mbox{ TeV}  \,\,\,\,\, \mbox{ 95\% C.L.} \,\,\, .
\end{equation}

The bounds from the $Z$ leptonic width and from the $\rho$ parameter 
are similar to the one from $G_F$, in that both arise only through the 
shifts in the gauge boson masses.  In the case of the leptonic width, we 
know that $\Gamma(\ell^+\ell^-)$ is proportional to 
$g^2 m_Z^{(ph)}/\cos^2\theta_w$, and that the $Z$ coupling is unaffected 
by the presence of extra dimensions, since the leptons are in the bulk.  
Hence, if we choose to express $\Gamma(\ell^+\ell^-)$ in terms of $G_F$ 
and $m_Z$ we find
\begin{equation}
\Gamma(\ell^+\ell^-) = \Gamma(\ell^+\ell^-)^{SM} \left(\frac{m_W^{(ph)2}}
{m_Z^{(ph)2} \cos^2\theta_w }\right) \,\,\, ,
\end{equation}
or using the result in Eq.~(\ref{eq:mzormw}),
\begin{equation}
\Gamma(\ell^+\ell^-) = \Gamma(\ell^+\ell^-)^{SM} \left[1+2\sin^2\theta_w
\sum_{n=1}\frac{m_Z^{(ph)2}}{n^2 M_c^2}\right]
\label{eq:gamlplm}
\end{equation}
Assuming the values $\Gamma=83.91\pm 0.10$ and $\Gamma^{SM}=84.00\pm 0.03$ 
given in the Review of Particle Physics~\cite{rpp98}, we find
\begin{equation}
M_c > 1.83 \mbox{ TeV}  \,\,\,\,\, \mbox{ 95\% C.L.} \,\,\, .
\end{equation}
The corrections to the $\rho$ parameter are also straightforward to
compute,
\begin{equation}
\rho=\rho^{SM}\left[1+2\sin^2\theta_w
\sum_{n=1}\frac{m_Z^{(ph)2}}{n^2 M_c^2}\right] \,\,\, .
\label{eq:rhoshift}
\end{equation}
Setting $\rho^{SM}= 1.0109 \pm 0.0006$~\cite{rpp98}, and computing $\rho$ 
from the measured $W$ and $Z$ masses, as well as the $\overline{MS}$ value 
of $\cos^2\theta_w$, $\rho = 1.0114 \pm 0.0023$, and we obtain the bound
\begin{equation}
M_c > 1.11 \mbox{ TeV}  \,\,\,\,\, \mbox{ 95\% C.L.} \,\,\, .
\end{equation}
This is superseded by the bounds that we have already obtained.

Atomic parity violation yields an even weaker bound.  The relevant
four-fermion operator $\overline{e} \gamma_\mu \gamma^5 e \overline{u}
\gamma^\mu u$ has the same dependence on $G_F$ and on the physical 
gauge boson masses as in the standard model, but an additional factor
of $1-2\sum_{n=1} m_Z^2/n^2M_c^2$ arising from the shift in the $Z$ boson 
coupling to quarks.  Thus, the weak charge is proportional to
$\rho (1-2\sum_{n=1} m_Z^2/n^2M_c^2)$, or using Eq.~(\ref{eq:rhoshift})
\begin{equation}
Q_W = Q_W^{SM} \left[1-2\cos^2\theta_w \sum_{n=1} 
\frac{m_Z^{(ph)2}}{n^2 M_c^2}\right] \,\,\, .
\end{equation}
Using the weak charge for Cesium $Q_W=-72.4\pm 0.84$, and
$Q_W^{SM}=-73.11 \pm 0.06$~\cite{rpp98}, we obtain the
bound
\begin{equation}
M_c > 802 \mbox{ GeV}  \,\,\,\,\, \mbox{ 95\% C.L.} \,\,\, .
\end{equation}

Finally, we consider the $Z$ hadronic width, $\Gamma(q\overline{q})$.  The 
difference between the derivation of $\Gamma(q\overline{q})$ and 
$\Gamma(\ell ^+ \ell^-)$ is the additional shift in $Z$-quark coupling 
by the factor given in Eq.~(\ref{eq:gphbrane}).  Thus, by comparison to 
Eq.~(\ref{eq:gamlplm}),
\begin{equation}
\Gamma(q\overline{q})=\Gamma(q\overline{q})^{SM}
(1+2\sin^2\theta_w  \sum_{n=1} 
\frac{m_Z^{(ph)2}}{n^2 M_c^2})(1-2\sum_{n=1} 
\frac{m_Z^{(ph)2}}{n^2 M_c^2})^2
\end{equation}
or 
\begin{equation}
\Gamma(q\overline{q})=\Gamma(q\overline{q})^{SM}
\left[1-2(2-\sin^2\theta_w)\sum_{n=1} 
\frac{m_Z^{(ph)2}}{n^2 M_c^2} \right] \,\,\, .
\end{equation}
Assuming the values $\Gamma(q\overline{q})=1.7432\pm 0.0023$~GeV, and
$\Gamma(q\overline{q})^{SM}=1.7433\pm 0.0016$~GeV, we obtain the
bound
\begin{equation}
M_c > 3.85 \mbox{ TeV}  \,\,\,\,\, \mbox{ 95\% C.L.} \,\,\, 
\end{equation}
which clearly supersedes all the other bounds, and places direct
production of KK modes outside of the reach of the Tevatron.
Although the lepton-gauge boson couplings were not affected in this model,
the choice of three-brane quarks and Higgs fields suggested by
gauge unification was sufficient to assure comparably stringent bounds.

Finally, we comment on the origin of the $\mu$ parameter in this
model.  Since both Higgs fields live on the three-brane, the 
$\mu$ parameter does not arise through compactification, but
is simply present as an allowed term in the Lagrangian.  Thus it
seems at first glance that the $\mu$ problem is no better than
in the minimal supersymmetric standard model (MSSM).  However, if
the fundamental cutoff of the theory $\Lambda_s$ (the string scale) 
coincides with the unification point, then we find 
$\Lambda_{s} \approx 56$~TeV for $M_c \approx 4$~TeV,
assuming the beta functions given in Ref.~\cite{carone1}.
To obtain a $\mu$ parameter below one TeV, we therefore would like
$\mu \approx 10^{-2} \Lambda_s$.  Such a suppression seems completely
natural from the point of view of horizontal flavor symmetries.
For example, in any flavor model in which the third generation
fields are trivial singlets, and in which the ratio of Higgs
vevs is of order unity, one might obtain the desired ratio
between bottom and top mass by assuming an additional approximate 
global symmetry under which only the down-type Higgs doublet transforms 
nontrivially.  Given the breaking of this symmetry in the quark Yukawa
interactions, one would then estimate that the $\mu$ parameter is of order 
$h_b \Lambda_s$, where $h_b$ is the bottom quark Yukawa coupling.
This is precisely of the desired magnitude.  There are presumably many 
ways in which a modest suppression factor such as this one can be obtained, 
so we will content ourselves with the observation that the $\mu$
problem seems less than problematic in this model given the greatly 
reduced ultraviolet cutoff of the theory.

\section{Bulk Generations} \label{sec:three}

The possibility of complete generations living in the bulk was
suggested in the work of Dienes, Dudas and Gherghetta~\cite{xdsm3} as 
a plausible variation on the minimal scenario.  Here we will consider 
the bounds on a model in which the first two generations live in the 
bulk, together with the gauge multiplets and one of the two MSSM Higgs 
fields.  This choice is preferred for a number of reasons.  First, we 
note that if only one of the first two generations lived in the bulk, 
then conservation of KK number would prevent the KK gauge bosons from 
coupling to zero-mode fields of that particular generation. The resulting 
violation of the Glashow-Iliopoulos-Miani (GIM) mechanism provides a 
much stronger constraint on the compactification scale, leading its 
decoupling from the weak scale or the scale of superparticle masses.  Let 
us perform some simple estimates.  Consider the upper two-by-two block of 
the Cabibbo-Kobayashi-Maskawa (CKM) matrix in Wolfenstein 
parameterization
\begin{equation}
V_{CKM}=U^{u\dagger}_L U^{d}_L \approx \left(\begin{array}{cc}
1 & \lambda \\
\lambda & 1 \\  \end{array} \right) \,\,\, ,
\end{equation}
where $\lambda\approx 0.2$ is the Cabibbo angle.  If we make a 
reasonable assumption that the Yukawa matrices and the biunitary 
matrices that diagonalize them are hierarchical in form, then we 
may parameterize
\begin{equation}
U_L^{u} = \left(\begin{array}{cc}
1 & a \lambda \\
-a \lambda & 1 \\
\end{array} \right)
\,\,\,\,\,\mbox{ and } \,\,\,\,\,
U_L^{d} = \left(\begin{array}{cc}
1 & b \lambda \\
-b \lambda & 1 \\
\end{array} \right)  \,\,\, ,
\end{equation}
with $b-a=1$.  Now consider the interaction between KK gluons and
quarks of the first two generations $q$, assuming that the
first generation lives in the bulk.  In the gauge basis, the
interaction vertex is given by
\begin{equation}
{\cal L}= \sum_{n=1} g_s \sqrt{2} \overline{q} G^{(n)}_\mu\gamma^\mu 
\left(\begin{array}{cc} 0 & 0 \\ 0 & 1 \end{array}\right) q
\end{equation}
where $G\equiv G^a T^a$ is the gluon field, and $g_s$ is the SU(3) 
gauge coupling.  This leads to flavor-changing interactions in the 
mass eigenstate basis, including
\begin{equation}
{\cal L} = \sum_{n=1} g_s \sqrt{2} \lambda \left[ 
(1-a) \overline{d}_L \not \! G^{(n)} s_L
+a \overline{u}_L \not \! G^{(n)} c_L + \cdots \right]
\end{equation}
Thus, we have at the very least the following two operators that 
contribute to $K$-$\overline{K}$ and $D$-$\overline{D}$ mixing
\begin{equation}
{\cal L}^{{\rm eff}} = 
\frac{1}{M_c^2}\frac{\pi^2}{6} g_s^2 \lambda^2
\left[(1-a)^2 (\overline{d}_L T^a \gamma^\mu s_L)^2
+a^2 (\overline{u}_L T^a \gamma^\mu c_L)^2
\right] \,\,\, ,
\end{equation}
where we have evaluated $\sum_{n=1} 1/n^2=\pi^2/6$ for $\delta=1$.  Using
the vacuum insertion approximation, it is straightforward to compute bounds 
from the splitting of the neutral meson mass eigenstates.
We find
\begin{equation}
M_c > 300 (1-a) \mbox{ TeV} \,\,\,\,\,\mbox{ and }\,\,\,\,\,
M_c> 120 a \mbox{ TeV}
\end{equation}
from $K$-$\overline{K}$ and $D$-$\overline{D}$ mixing, respectively.
This implies an absolute lower bound of $M_c > 85$~TeV (for
$a\simeq 0.71$) which places the lowest KK mode well above the 
electroweak scale, and beyond the reach of any proposed 
collider experiment.

On the other hand, the possibility that the third generation is
distinguished by its bulk/brane assignment is far less constrained.
The relevant term in the effective Lagrangian, assuming CKM-like
mixing angles is
\begin{equation}
{\cal L}^{eff} = \frac{1}{M_c^2}\frac{\pi^2}{6} g_s^2 \lambda^8 c
(\overline{b}_L T^a \gamma^\mu d_L)^2
\label{eq:bbop}
\end{equation}
and yields a bound of $M_c > 1.33 c$~TeV, where $c$ is an operator
coefficient.  This bound can be evaded, however, if the third generation 
CKM angles originate only from rotations on the left-handed up quarks.  
Notice that even the modest choice of $c=1/2$ renders the bound from 
Eq.~(\ref{eq:bbop}) weaker than the typical bounds we encountered in 
the previous section.  Other third generation flavor-changing processes
may be interesting as signals for this type of model, but at present 
do not provide any meaningful constraints~\cite{burd}. 

The observations above hold true if either the third generation is in 
the bulk and the first two generations are on the three-brane, or vice 
versa.  However, the first choice seems disfavored by the largeness of 
the top quark Yukawa coupling.  The difficulty originates from the
rescalings that one must perform to relate Yukawa couplings in the 
$4+\delta$ dimensional theory to the Yukawa couplings we know and love.   
For example, in the case where the Higgs and the top quark both live
in the bulk, then a Yukawa coupling of the higher dimensional
theory has mass dimension $-\delta/2$.  By a naturalness argument,
we might expect this dimensionful coupling to be of the same order as the 
cut off of the theory, so that  $h \sim h_0/\Lambda_s^{\delta/2}$, where 
$h$ is the Yukawa coupling, $\Lambda_s$ is the string scale, and $h_0$ is 
dimensionless and of order unity.  However, when one derives the effective 
four dimensional Lagrangian in terms of 4D fields with canonical mass 
dimension, one finds that
\begin{equation}
h_{4D} = h_0 \left(\frac{\Lambda_c}{\Lambda_s}\right)^{\delta/2}  \,\,\, ,
\label{eq:rescale}
\end{equation}
where $\Lambda_c=1 / (2 \pi R)$. We find the same result for bulk Higgs and 
brane fermions, while in the case of bulk fermions and brane Higgs the 
exponent changes from $\delta/2$ to $\delta$.  In light of this
result, an order one top quark Yukawa coupling suggests that all of 
the associated fields live on the three-brane.   This leads us to the
scenario of interest, in which only the first two generations, gauge 
fields and down-type Higgs live in the bulk.

With all the gauge fields and one Higgs in the bulk, the $W$ and $Z$
mass eigenvalues, as well as the shift in their couplings to three-brane
fermions are the same as those presented in Ref.~\cite{ewbds2}:
\begin{equation}
m^{(ph)2}_{W,Z} = m^{(ph)2}_{W,Z} \left[1-2\sin^4\beta \sum_{n=1}
\frac{m^2_{W,Z}}{n^2 M_c^2}\right]
\end{equation}
\begin{equation}
g^{(ph)}_{brane}=g \left[1-2\sin^2\beta\sum_{n=1}
\frac{m^2_{W,Z}}{n^2 M_c^2}\right]
\end{equation}
where $\tan\beta$ is the ratio of brane to bulk Higgs vevs.  As in the 
bulk lepton scenario, however, conservation of KK number prevents couplings 
between the KK gauge bosons and {\em any} zero mode of the first two 
generations. As we will see, this has a significant impact on the form of 
the electroweak constraints.  In the case of $G_F$, $\Gamma(e^+e^-)$, and 
$\rho$, the analysis differs only trivially from that described in 
Section~\ref{sec:two}, so we will simply state the results:
\begin{equation}
G_F = G_F^{SM} \left[1 - 2 \sin^4\beta \sum_{n=1}
\frac{m^{(ph)2}_W}{n^2 M_c^2}\right]
\end{equation}
\begin{equation}
\Gamma(e^+e^-) = \Gamma(e^+e^-)^{SM} \left[1+2\sin^2\theta_w \sin^4\beta
\sum_{n=1} \frac{m^{(ph)2}_Z}{n^2 M_c^2} \right]
\end{equation}
\begin{equation}
\rho = \rho^{SM} \left[1+2 \sin^2\theta_w\sin^4\beta
\sum_{n=1} \frac{m^{(ph)2}_Z}{n^2 M_c^2} \right] \,\,\, .
\label{eq:newrho}
\end{equation}
Note that for the sake of simplicity we have chosen to study $\Gamma(e^+e^-)$ 
rather than the full leptonic width, which has a more complicated form given 
the differing $Z$ coupling to third generation leptons; we expect the bounds 
to be similar.  The correction to the weak charge $Q_W$ does not follow
directly from the result in Section~\ref{sec:two}, since in this 
case $Q_W$ is only altered by the shift in gauge boson masses:
\begin{equation}
Q_W = Q_W^{SM} \left[1+2 \sin^2\theta_w \sin^4 \beta
\sum_{n=1} \frac{m^{(ph)2}_Z}{n^2 M_c^2} \right] \,\,\, .
\label{eq:newqw}
\end{equation}
Eq.~(\ref{eq:newqw}) reflects the fact that $Q_W \propto \rho$ as given
in Eq.~(\ref{eq:newrho}).  Finally, we consider the $Z$ decay width to 
$\overline{b} b$.  We find
\begin{equation}
\Gamma(b\overline{b}) = \Gamma(b\overline{b})^{SM} \left[
1-2\sin^2\beta(2-\sin^2\beta\sin^2\theta_w) \sum_{n=1}
\frac{m^{(ph)2}_Z}{n^2 M_c^2} \right] \,\,\, .
\end{equation}
For $G_F$, $\rho$ and $Q_W$, we use the experimental and standard model values
stated in Section~\ref{sec:two}.  In addition, we assume $\Gamma(e^+e^-)=
83.8134 \pm 0.3085$~MeV, $\Gamma(e^+e^-)^{SM}=84.01 \pm 0.05$~MeV, 
$\Gamma(b\overline{b})=0.3783 \pm 0.0016$, and 
$\Gamma(b\overline{b})^{SM}=0.3762 \pm 0.0004$, 
computed from branching fractions given in the Review of Particle 
Physics~\cite{rpp98}. We then obtain the bounds shown in Figure~1.  
\begin{figure}  
\centerline{ \epsfxsize=4.5 in \epsfbox{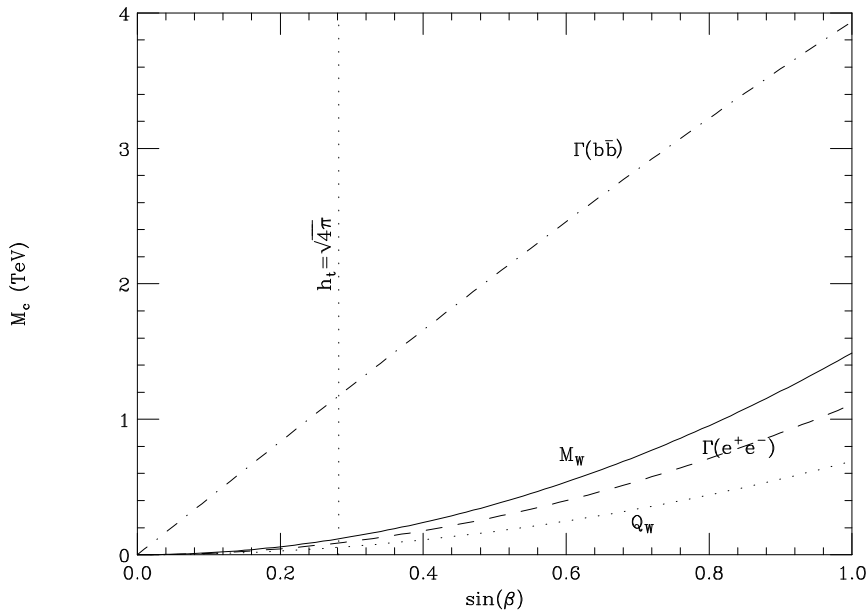}  }
\vglue .3 in
\caption{Bounds on the bulk generations scenario.  The $h_t$ line
indicates where the top quark Yukawa coupling becomes nonperturbative.  
The bound from $\rho$ is indistinguishable from the $\Gamma(e^+e^-)$ result.}
\end{figure}

The most significant bound shown is the one from $Z\rightarrow b \overline{b}$
and is as high as $\sim 3.9$~TeV in the case where $\sin\beta\approx 1$.  
However, an interesting feature that these bounds share is that they vanish 
(at least to the order we are working) in the opposite limit, 
$\sin\beta \rightarrow 0$.  This result is easy to understand 
qualitatively:  Since the first two generations live in the bulk, corrections 
to electroweak observables that arise from KK exchange diagrams 
(which are independent of $\sin\beta$) are suppressed.  The remaining 
contributions arise from shifts in the gauge boson masses and couplings, 
and are all proportional to powers of $\sin\beta$ since the gauge boson 
mixing depends on the magnitude of the brane Higgs vev.  

How small then can we reasonably take $\sin\beta$?  In the MSSM, this 
issue is normally settled by consideration of the perturbativity of the 
top quark Yukawa coupling, renormalized up to high energy scales.  As 
pointed out in Ref.~\cite{xdsm3}, the Yukawa couplings in the minimal 
scenario are actually driven to weaker values above the 
compactification scale, and this observation should carry over to the 
top quark Yukawa coupling in the model of interest here.  However, 
even taking the top quark Yukawa to be as large as $\sim 3$ at the weak 
scale, we can only bring the tightest bound from $\Gamma(b\overline{b})$ 
down to $\sim 1$~TeV.  Conservation of KK number for the light generations 
implies that KK modes are primarily pair produced in this model, so 
direct production channels would remain out of the reach of the Tevatron.  
A possible, albeit speculative, loophole is that $\sin\beta$ could be 
reduced much further if the top quark mass were generated largely via 
other (strong) dynamics.  This is interesting in that it has been 
suggested that models of dynamical electroweak symmetry breaking 
involving top condensation may have a natural origin in an extra dimensional 
framework~\cite{dobrescu}.  In this case, the results described here 
would suggest that KK excitations could be brought down to sub-TeV 
energy scales, without running afoul of precision electroweak 
constraints\footnote{The bulk generation scenario described here
would also provide a natural means of restricting extra-dimensional
strong dynamics to the third generation.}.

\section{An SU(2)-brane scenario} \label{sec:four}

In this section, we consider an even more unconventional possibility, 
that different factors of the electroweak gauge group have different 
bulk/brane assignments.  In Refs.~\cite{carone1,xdgu2} it was pointed 
out that assigning the SU(2) gauge multiplet to the three-brane is
consistent with gauge unification, at least from a bottom-up point of view, 
if the other gauge multiplets, the right-handed leptons, and one generation 
of right-handed up and down quarks are assigned to the bulk.  However, it 
was noted in  Ref.~\cite{kak} that a unified boundary condition for 
the couplings at the high scale may not be expected generically if the 
gauge groups are differentiated in this way.  Here we will simply allow 
for the possibility of a unified boundary condition, and focus on 
low-energy phenomenology.  As we suggested in the introduction, any model
whose electroweak sector is extended by allowing only the hypercharge
multiplet to propagate in the bulk, might be thought of as possessing
a minimal extra-dimensional $Z'$, the first KK mode of the hypercharge
gauge boson.  While we will assume the bulk/brane assignments of the
matter fields stated in Ref.~\cite{carone1} and above, we will describe 
to what extent our results carry over to any model of this type.

If we assign SU(2) to the three-brane, then we must do the same for 
all the SU(2) doublet fields.  With two three-brane Higgs doublets, we will 
have, as in Section~\ref{sec:two}, mixing between the zero-mode and KK 
gauge bosons that is independent of $\sin\beta$.  However, the form of the 
mixing matrix is quite different.  The neutral gauge boson mass terms in 
the four dimensional Lagrangian are given by
\[
{\cal L}=\frac{1}{4} (v_1^2+v_2^2) \left[
g^2 W_3^\mu W_\mu^3 - 2 g g' W_\mu^3(B^{\mu (0)} + \sqrt{2} \sum_{n=1} 
B^{\mu (n)}) \right.
\]
\begin{equation}
\left. +{g'}^2(B^{\mu (0)} + \sqrt{2} \sum_{n=1} B^{\mu (n)})^2
\right] 
+ \sum_{n=1}\frac{1}{2}\frac{n^2}{R^2} B^{\mu (n)} B_{\mu}^{(n)} \,\,\, ,
\end{equation}
where $W$ and $B$ are the SU(2) and U(1) gauge fields, respectively.
Rewriting the zero-mode fields in terms of conventionally defined photon 
and $Z$ fields, we then obtain a mixing matrix between the $Z$ boson, and 
the KK excitations of the hypercharge gauge field $B$.  In the 
basis $(Z, B^{(1)}, B^{(2)} \cdots)$, we obtain the mass matrix
\begin{equation}
\left(\begin{array}{cccc}
m^2_Z & -\sqrt{2} s_w m_Z^2 & -\sqrt{2} s_w m_Z^2 & \cdots \\
-\sqrt{2} s_w m_Z^2 & M_c^2 & 2 s_w^2 m_Z^2 & \cdots \\
-\sqrt{2} s_w m_Z^2 & 2 s_w^2 m_Z^2 & (2M_c)^2 & \\
\vdots & \vdots & & \ddots \\
\end{array}\right) \,\,\, ,
\end{equation}
where we have written $\sin\theta_w$ as $s_w$ for shorthand.
Working again to lowest order in $m_Z^2/M_c^2$, this
matrix is diagonalized by the rotation
\begin{equation}
R=
\left(\begin{array}{ccccc}
1 & \theta_1 & \theta_2 & \cdots & \theta_n  \\
-\theta_1 & 1 & \theta_{12} & \cdots & \theta_{1n} \\
-\theta_2 & -\theta_{12} & 1 & \cdots & \theta_{2n}  \\
\vdots & \vdots  &  & \ddots & \\
-\theta_n & -\theta_{1n} & & & \theta_{nn} \\
\end{array}\right)   \,\,\,\,\,\,\,\,\,\, 
\begin{array}{c}
\theta_n=\frac{\sqrt{2} s_w m_Z^2}{n^2 M_c^2} \\ 
\\
\theta_{ij}=-\frac{1}{|i^2-j^2|}\frac{2 s_w^2 m_Z^2}{M_c^2} \\
\end{array}
\end{equation}
yielding the lightest eigenvalue
\begin{equation}
m_Z^{(ph)2}=m_Z^2 \left[1-2\sin^2\theta_w \sum_{n=1} \frac{m_Z^2}{(nM_c)^2}
\right] \,\,\, .
\end{equation}
The shift in the $Z$ coupling to brane fermions is slightly more
complicated then in the other models we have considered.  Writing
the vertex in terms of the third component of isospin $T^3$ and
hypercharge $Y$, we obtain
\begin{equation}
-\frac{e}{s_w c_w} \left[c_w^2 T^3 - s_w^2 Y (1+2\sum_{n=1}
\frac{m_Z^2}{n^2 M_c^2})\right] \,\,\, ,
\end{equation}
which reflects the fact that only the U(1) gauge field has
KK excitations.  The $Z$ coupling to zero-mode bulk fermions remains the
same as in the standard model.  We are now ready to determine the
electroweak constraints.

In this scenario, the $W$ has no KK excitations, and its mass eigenvalue 
remains unaffected by the presence of extra dimensions.  However, the shift 
in the $Z$ mass affects $G_F$ through the on-shell definition 
of $\sin^2\theta$:
\begin{equation}
G_F=G_F^{SM} \left[1-2\sin^2\theta_w \frac{m_W^{(ph)2}}
{m_Z^{(ph)2}-m_W^{(ph)2}}\sum_{n=1} \frac{m_Z^{(ph)2}}{n^2 M_c^2}
\right] \,\,\, .
\end{equation}
From this we obtain the bound 
\begin{equation}
M_c > 1.52\mbox{ TeV}  \,\,\,\,\, \mbox{ 95\% C.L.} \,\,\, .
\label{eq:mib}
\end{equation}
The $Z$ leptonic width on the other hand receives corrections from 
two sources in this scenario: the shift in the $Z$ mass, and the altered 
coupling to left-handed (three-brane) leptons.  If we write the 
$Z\overline{e}e$ coupling in terms of its vector and axial vector 
components, $g_V$ and $g_A$, then the shift in the $Z$ coupling to the 
left-handed component gives us a correction
\begin{equation}
\Delta g_V = \Delta g_A \equiv \Delta g = -\frac{e}{2 s_w c_w} (s_w^2
\sum_{n=1} \frac{m_Z^2}{n^2 M_c^2})
\end{equation}
Thus the two effects described above lead to the form
\begin{equation}
\Gamma(\ell^+\ell^-) = \Gamma(\ell^+\ell^-)^{SM}
\left(1+2\Delta g \frac{g_V+g_A}{g_V^2+g_A^2}\right)
\left(\frac{m_W^{(ph)2}}{m_Z^{(ph)2} \cos^2\theta_w}\right)
\end{equation}
or after some algebra
\begin{equation}
\Gamma(\ell^+\ell^-) = \Gamma(\ell^+\ell^-)^{SM}
\left[1-2\sin^2\theta_w 
\left(\frac{1-8\sin^4\theta_w}{1-4\sin^2\theta_w+8\sin^4\theta_w}
\right)\sum_{n=1}\frac{m_Z^{(ph)2}}{n^2M_c^2} \right]  \,\,\, .
\label{eq:gee}
\end{equation}
This gives us a bound comparable to (\ref{eq:mib})
\begin{equation}
M_c > 1.53\mbox{ TeV}  \,\,\,\,\, \mbox{ 95\% C.L.} \,\,\, .
\label{eq:geebd}
\end{equation}

Given the breaking of custodial isospin in this scenario, one
might expect a significant bound from the $\rho$ parameter.
Since the ratio of $g_V/g_A$ is shifted away from the standard model 
value, we must take into account the effect on $\sin^2\theta^{eff}$
(from which we determine the corresponding $\overline{MS}$ value)
as well the shift in the $Z$ mass in computing $\rho$.  We find
\begin{equation}
\sin^2\theta^{eff} = \sin^2 \theta_w \left[1+2\sin^2\theta
\sum_{n=1} \frac{m_Z^{(ph)2}}{n^2M_c^2} \right] \,\,\, ,
\end{equation}
and
\begin{equation}
\rho = \rho^{SM} \left[1+2\tan^2\theta_w
\sum_{n=1} \frac{m_Z^{(ph)2}}{n^2 M_c^2} \right] \,\,\, 
\end{equation}
from which we conclude
\begin{equation}
M_c > 1.26\mbox{ TeV}  \,\,\,\,\, \mbox{ 95\% C.L.} \,\,\, .
\end{equation}
While this is a stronger bound than we obtained from
consideration of the $\rho$ parameter in the bulk lepton
and bulk generation scenarios, it does not supersede
Eq.~(\ref{eq:geebd}).  Unlike the other scenarios, we obtain
a competitive bound from atomic parity violation,
\begin{equation}
Q_W = Q_W^{SM} \left[1+\frac{10 \sin^2\theta_w}
{3-8\sin^2\theta_w} \sum_{n=1}\frac{m_Z^{(ph)2}}{n^2M_c^2} \right] \,\,\, ,
\end{equation}
yielding
\begin{equation}
M_c > 1.44\mbox{ TeV}  \,\,\,\,\, \mbox{ 95\% C.L.} \,\,\, ,
\end{equation}
assuming three-brane quarks.  If we allow ourselves the freedom
to stray from the bulk/brane assignments of the matter fields
given in Ref.~\cite{carone1}, then we would expect the bounds
to vary in a model-dependent way.  The exception, however, is the 
bound from $G_F$: in any variant of this model, the left-handed fields 
are again located on the three-brane, so that the form of $G_F$ as
determined in muon decay remains unchanged.  Since the gauge boson
mass matrix is also the same, the bound $M(Z'_{XD})>1.52$~TeV is 
model independent.
 
\section{Conclusions}

In each of the extended models considered in this paper we have
found that electroweak constraints lead to typical bounds of order a 
few TeV.  In the bulk lepton scenario, KK excitations of the gauge fields 
cannot couple at lowest order to the lepton zero modes.  Nevertheless, 
the fact that gauge unification required that we place both Higgs fields 
on the three-brane yielded unavoidable tree-level $Z$-KK mixing, resulting 
in bounds as large as $3.85$~TeV.  Such mixing was also inherent to the 
SU(2)-brane scenario, forcing $M_c > 1.52$~TeV, from consideration of 
the Fermi constant.  Given the typical results of recent collider
studies~\cite{ewbds3}, we conclude that both the bulk lepton and 
the SU(2)-brane scenarios are outside the reach of the Tevatron for direct 
production of KK states, but nonetheless could be discovered at the LHC.  
The bulk generation scenario is interesting in that the lowest 
order bounds weaken monotonically as $\sin\beta$ is decreased, the same
limit in which the brane Higgs vev vanishes.  Since this is the Higgs field 
that is responsible for giving the top quark its mass, we found that
the requirement of perturbativity of the top quark Yukawa coupling
leads to bounds  of the same order as those in the other two scenarios. 
However, we noted that in models where the top quark mass has an
additional dynamical component (which have been suggested in the extra 
dimensional context~\cite{dobrescu}) that $\sin\beta$ could be reduced 
and the electroweak bounds weakened, allowing the possibility of sub-TeV KK 
excitations.  Finally, we point out that the extended models considered 
here have collider signatures that differ noticeably from the minimal 
scenario, ranging from leptophobic $W'$ bosons in the bulk lepton scenario, 
to flavor-changing neutral current KK interactions involving third generation 
fields in the bulk generation scenario.  Thus, some aspects of the collider 
phenomenology of these models may be worthy of further study.

{\samepage
\begin{center}
{\bf Acknowledgments}
\end{center}
CC thanks the National Science Foundation for support under Grant No.\
PHY-9800741 and PHY-9900657.}


\end{document}